\documentstyle[multicol,aps,epsf]{revtex}
\begin{document}
\title{Numerical Study of the Vortex Phase Diagram
Using the Bose Model \\
in STLS approximation}
\author{Bahman Davoudi$^a$ and Mohammad Kohandel$^{a,b}$}
\address{$^a$ Institute for Studies in Theoretical Physics and Mathematics,
Tehran 19395-5531, Iran \\
$^b$ Institute for Advanced Studies in Basic Sciences,
Zanjan 45195-159, Iran}
\date{\today}
\maketitle
\begin{abstract}
We study the phase diagram of the flux lines using the mapping
to 2D bosons in the self-consistent-field approximation of Singwi,
Tosi, Land, and Sjolander (STLS). 
The pair correlation function, static structure factor, 
interaction energy, and spectrum of the excited energies are 
calculated over a wide range of the parameters
in this approximation. These quantities are used for studying 
the melting transition from the Abrikosov lattice into
the entangled vortex liquid. The resulting
$B-T$ phase diagram is in good agreement with the 
known estimates for the vortex lattice melting and the
Monte Carlo simulations.
We also discuss the effect of van der Waals interaction,
induced by thermal fluctuations,
together with the repulsion potential on the phase diagram. 
\end{abstract}
\begin{multicols}{2}

\section{Introduction}
The phases of flux lines (FLs) in high temperature superconductors
are the subject of many current experimental and theoretical 
investigations \cite{Blatter1,Brandt}. 
In the classical (Abrikosov) picture of type-II superconductors, 
FLs penetrate a clean material for fields $H>H_{c1}$, to form a 
triangular solid lattice. 
This mean field phase diagram is modified by inclusion of 
the strong thermal fluctuations and play an important role 
in high-$T_c$ 
materials \cite{Blatter1,Brandt,Larkin,Nelson1,Fisher,MK1}.
The presence of various forms of disorder  (such as point 
or columnar disorder), also affects on the mean field phase diagram.
In this paper we typically focus on the phase diagram of the
pure materials, however, there are many interesting 
experimental and theoretical 
works describing the effect of both thermal fluctuations and 
disorder on the phase diagram \cite{Nelson,Tauber,Kardar,MK2}. 

It is well known that the thermal fluctuations lead to melting
of the vortex lattice and appearance of a vortex liquid.
It has now become possible to observe the melting transition
experimentally. The indirect observations \cite{Safar,Char} 
based on the resistivity
measurements, and recent experiments \cite{Zeldov,Welp,Schil,Roulin} 
based on the measuring a jump
in the magnetization or the latent heat show a first-order vortex
lattice melting transition.
On the other hand, vortex lattice melting has been studied theoretically
using various approximations. Early works using the 
renormalization group \cite{Brezin} or density functional theory
\cite{Seng} have indicated a first-order transition. 
Elastic theory combined with the Lindemann criterion 
produce a melting line in good agreement
with the experimental observations \cite{Blatter2}.
There are also a large interest in numerical simulations
for studying this problem. An interesting work in this direction
is done by Nordborg and Blatter \cite{Blatter3} which present an extensive 
numerical study of the vortex matter using the mapping to 2D bosons
and path-integral Monte Carlo simulations.

It was suggested by Nelson that the vortex system is equivalent 
to a system of interacting bosons in two dimensions (Bose model)
\cite{Nelson1,Nelson}.
This mapping predicts a melting transition into an entangled vortex
liquid. 
Therefore, the problem of a vortex system maps to a system
of $N$ bosons in two dimensions interacting through the 
potential $V(r)=g^2 K_0(r/\lambda)$, where $K_0$ is the
modified Bessel function, $\lambda$ is the London penetration
depth, and $g^2$ is a constant that scales the energy of
interaction. In the language of the vortices, this potential
comes from the interaction between vortices in the London theory, and 
$g$ is related to the elastic moduli of the vortex lattice.
The Bose model differs from the real vortex 
system (see next section), however it still contains the main
part of the interaction \cite{Blatter3}. Hence, it
would be reasonable and interesting
to use this model for describing the properties of the vortex 
phase diagram. 
More recently \cite{Blatter3}, the Bose model has 
been used in a numerical study of the vortex matter.
Some physical quantities such as structure factor
and superfluid density in different temperatures are given addressed,
and the first order vortex lattice melting transition 
into an entangled vortex liquid is approved by numerical simulations. 
In the language of
the boson system, this transition is related to the quantum 
phase transition from a Wigner crystal to a superfluid.

The Bose model idea also allows for using the many body techniques.
In this work, we apply the self-consistent-field approximation of Singwi,
Tosi, Land, and Sjolander (STLS) \cite{STLS}, and calculate the 
static structure factor, pair correlation function, 
interaction energy and the spectrum of the excited energies 
for different magnetic field strengths and temperatures.
The STLS approximation has
originally been proposed for describing a degenerate electron gas
and has been used successfully to study a variety of other
systems too \cite{Singwi,Moudgil}. 
In the STLS theory, the correlation effects are incorporated 
through a static local-field correction, which is obtained
numerically in a self-consistent way. 
We find numerical results for the static structure factor
$S(q)$ over a wide range of the parameters. From the calculation of $S(q)$,
we present the results for the pair 
correlation function, the interaction energy and 
spectrum of the excited energies. 
Different behaviors of these quantities may be used for 
studying the phases of vortex lines.

It is well known that the oscillatory behavior in $g(r)$ 
is a signature of the solid phase.
Therefore, the phase transition can be detected 
by looking at the behavior of the $g(r)$ in a fixed temperature
(magnetic field) but varying magnetic field (temperature).
The solid-liquid transition can also be observed using the 
static structure factor. Disappearing of the peaks in the structure 
factor resembles the onset of the phase transition.   
Hence, using the behaviors of the pair correlation function  
and static structure factor, the phase diagram can be explained
qualitatively, however, it is not possible
to determine precisely the melting transition temperature. 

One of the transition signatures 
is the appearance of a special $q$ on which the
spectrum of the excited energies vanishes.
So we have numerically investigated the dispersion
relation of the excited energies as a good
quantity for the estimation of the $B-T$ 
phase diagram and the melting temperature.
Quantitatively our results
for the excited energies are compatible with the expected results of 
the phase diagram \cite{Blatter1} and 
recent Monte Carlo simulations \cite{Blatter3}.

On the other hand, the direction dependent interaction for
real vortices has very interesting consequences and predicts
a van der Waals interaction even for straight vortices \cite{Blatter4}.
It is shown \cite{Blatter4,Volmer1} that in the decoupled limit, 
$\gamma\rightarrow 0$ (where $\gamma$ is the anisotropy parameter), 
the van der Waals attraction is proportional to $1/R^4$.
This attractive interaction with entropic repulsion has very 
important outcomes for the low-field phase diagram of the
anisotropic superconductors \cite{Blatter4,Volmer1}.
Recently Volmer and Schwartz \cite{Volmer2} introduced a new variational
approach to consider the effects of van der Waals attraction
and repulsive interaction on the field phase diagram.
We also consider the same model in the STLS approximation
for studying the combined effects of the repulsive and attractive 
potentials on the 
solid and liquid phases of the pure anisotropic or layered superconductors.

The rest of this paper is organized as follows.
In Sec.~II, we shortly review the Bose model and discuss its 
applicability to the vortex system.
The STLS approximation is briefly discussed in Sec.~III. 
The numerical results for the repulsion interaction is presented
and discussed in the Sec.~IV. The results for the van der Waals
interaction and its
consequences in the phase diagram is given in Sec.~V, and
finally the conclusions appear in Sec.~VI.

\section{Bose model}
In the Feynman path-integral picture \cite{Feynman}, the 
system of $N$ interacting bosons in two dimensions is  
described by the action
\begin{equation}\label{actionS}
\frac{S}{h}=\frac{1}{h}\int_0^{h/T} d\tau \left\{\sum_i \frac{M}{2}
\left(\frac{d\vec R_i}{d\tau}\right)^2+\sum_{i<j} g^2 
K_0\left(\frac{R_{ij}}{\lambda}\right)\right\},
\end{equation}
where $\vec R_i(\tau)$ is a two dimensional vector representing
the positions of the bosons, and $T$ is the temperature of the
system. In the Schrodinger picture the above action is
equivalent to the two-dimensional Schrodinger equation,
\begin{equation}\label{shrod}
\left[-\sum_i\frac{\nabla_i^2}{2M}
+\sum_{i,j}V(R_{ij})\right]\psi_0=E_0\psi_0,
\end{equation}
where the potential $V(R_{ij})$ is proportional to the modified 
Bessel function. 

It was pointed out by Nelson \cite{Nelson1,Nelson} that 
the above action can be also 
interpreted as the London free energy for a system of vortex lines.
The London free energy for a system of interacting
vortices for a sample of length $L_z$ is given by,
\begin{equation}\label{actionF}
\frac{{\cal F}}{T}=\frac{1}{T}\int_0^{L_z} dz \left\{\sum_i 
\frac{\varepsilon_l}{2}
\left(\frac{d\vec R_i}{dz}\right)^2+\sum_{i<j} 2\varepsilon_0 
K_0\left(\frac{R_{ij}}{\lambda}\right)\right\},
\end{equation}
where $\varepsilon_l\approx \gamma^2\varepsilon_0 
a_0/(2\sqrt{\pi}\xi)$, $\varepsilon_0=(\Phi_0/4\pi\lambda)^2$,
$\Phi_0$ is the quantum flux, $\xi$ is coherence length, and $a_0$ is  
the lattice spacing.
Comparing this functional free energy, 
which is referred  
as Bose model for the vortex system, 
with the action (\ref{actionS}) shows 
the relationship between the parameters 
of the 2D bosons and vortices.

The modified Bessel function $K_0(R/\lambda)$ describes
a screened logarithmic interaction, $K_0(x) \sim -ln(x)$ 
for $x\rightarrow 0$, and $K_0(x) \sim x^{-1/2} e^{-x}$ for
$x\rightarrow \infty$. Thus, the London penetration depth
defines the interaction range. Note that according to the
two-fluid model \cite{Nelson,Tauber}, the penetration depth
diverges at zero-field transition temperature $T_c$, and 
therefore the interaction range becomes considerably
longer upon approaching $T_c$.

As it is discussed in details in the Ref. \cite{Blatter3},
in spite of the fact that the Bose model differs from the real 
vortex system in the choice of boundary conditions, 
the linearalization leading to the elastic term, and
the retarded interaction, contain
the main parts of the interaction between vortices and one expects
that the results be in a rough quantitative agreement with those of 
real systems.

\section{STLS approximation}
In this section we shortly review the principal equations 
of the STLS approximation \cite{Moudgil}. 
These equations show the relation between the response function, 
static structure factor, and the local field correction.

In the STLS approximation, the response function 
can be expressed as \cite{Singwi},
\begin{equation}\label{resf}
\chi(q,\omega)=
\frac{\chi_0(q,\omega)}{\left[1-\psi(q)\chi_0(q,\omega)\right]}.
\end{equation}
In this equation $\chi_0$ is the response function of the free 
Bose gas and $\psi$ is the effective potential given by 
$\psi(q)=v(q)(1-G(q))$, where $v(q)$ is the Fourier transformation 
of the bare potential, and $G(q)$ is local field correction. The 
STLS local field correction is 
\begin{equation}\label{LFC}
G(q)=-\frac{1}{n}\int \frac{d{\bf q}'}
{(2\pi)^2}\frac{({\bf q}.{\bf q}')}{q^2}
\frac{v(q')}{v(q)}(S(|{\bf q}-{\bf q}'|)-1),
\end{equation}
where $n$ is the density, and $S(q)$ is the static structure factor 
and is related to response function as ($\hbar=1$),
\begin{equation}\label{S1}
S(q)=-\frac{1}{(n\pi)}\int_0^\infty d\omega Im(\chi(q,\omega)).
\end{equation}
For a noninteracting two dimension Bose gas the free response function 
is given by,
\begin{equation}\label{chi0}
\chi_0(q,\omega)=\frac{2n\epsilon(q)}
{\left[(\omega+i\eta)^2-\epsilon(q)^2\right]},
\end{equation}
where $\epsilon(q)=q^2/(2m)$ is the free particle energy, and
$\eta$ is a positive infinitesimal quantity.

Using the Eqn.~(\ref{chi0}), one can calculate the integral 
Eqn.~(\ref{S1}) analytically which leads to the following result 
for the static structure factor 
\begin{equation}\label{S2}
S(q)=\frac{1}{\left[1+2n\psi(q)/\epsilon(q)\right]^{(1/2)}}.
\end{equation}
The Eqs.~(\ref{LFC}) and~(\ref{S2})
should be solved numerically for obtaining the $S(q)$ self consistently. 
From the knowledge of the structure factor, the pair correlation
function $g(r)$ is calculated as
\begin{equation}\label{gr}
g(r)=1+\frac{1}{n}\int\frac{d\bf q}{(2\pi)^2}e^{i{\bf q}.{\bf r}}
\left[S(q)-1\right].
\end{equation}

The interaction energy is also related to the structure factor as,
\begin{equation}\label{Ec}
E_{int}=\frac{1}{4\pi}\int_0^1 \frac{d\lambda}{\lambda}\int  
v_{\lambda}(q)(S_{\lambda}(q)-1)qdq.
\end{equation}
where $v_{\lambda}(q)=\lambda v(q)$ and $S_{\lambda}(q)$ is its
related static structure factor.

One can also compute the excited energy using the poles
of Eqn.~(\ref{resf}), leading to
\begin{equation}\label{omegaq}
\omega(q)=\sqrt{\epsilon^2(q)+2n\epsilon(q)v(q)(1-G(q))}.
\end{equation}

In the next section, we present numerical results for various 
quantities of interest.

\section{Numerical results}
In this section we present the results of the numerical calculations
for the interesting physical quantities.
We numerically solve 
the set of Eqs.~(\ref{LFC}) and~(\ref{S2})
with the repulsion potential defined in 
Eqn.~(\ref{actionF}), and find the static structure factor.
The calculations are done for different values of the two parameters 
$m$ and $r_s$.
For the boson system $m$ can be considered as the mass of particles 
(in the unit of $\hbar=1$), and for
vortex lines $m$ is related to the temperature as $m=\varepsilon_0
\varepsilon_1\lambda^2/T^2$. 
Substituting
the numerical values for parameters as 
$\varepsilon_0=50K/A$, $\gamma=100$, $\lambda=1000A$,
the equivalent temperature will be fixed as $T\approx 500/\sqrt{m} (K)$.
On the other hand,
$r_s$ is the inverse of density corresponding to the particle density 
for boson system, or the density of FLs in the vortex matter. 
The mentioned relationship is expressed as $B=\phi_0/(\pi r_s^2\lambda^2)$,
or $B\approx 0.06/r_s^2$ (Tesla).

One of our numerical results which was developed by iterating the 
Eqs.~(\ref{LFC}) and~(\ref{S2}) is 
the behavior of the static structure factor which is shown in Fig.~1
for a fixed $r_s=0.6$ ($B\approx 0.17$(Tesla)) and different $m$'s (temperatures).

\begin{figure}\label{Fig1}
\epsfxsize=8truecm 
\centerline{\epsfbox{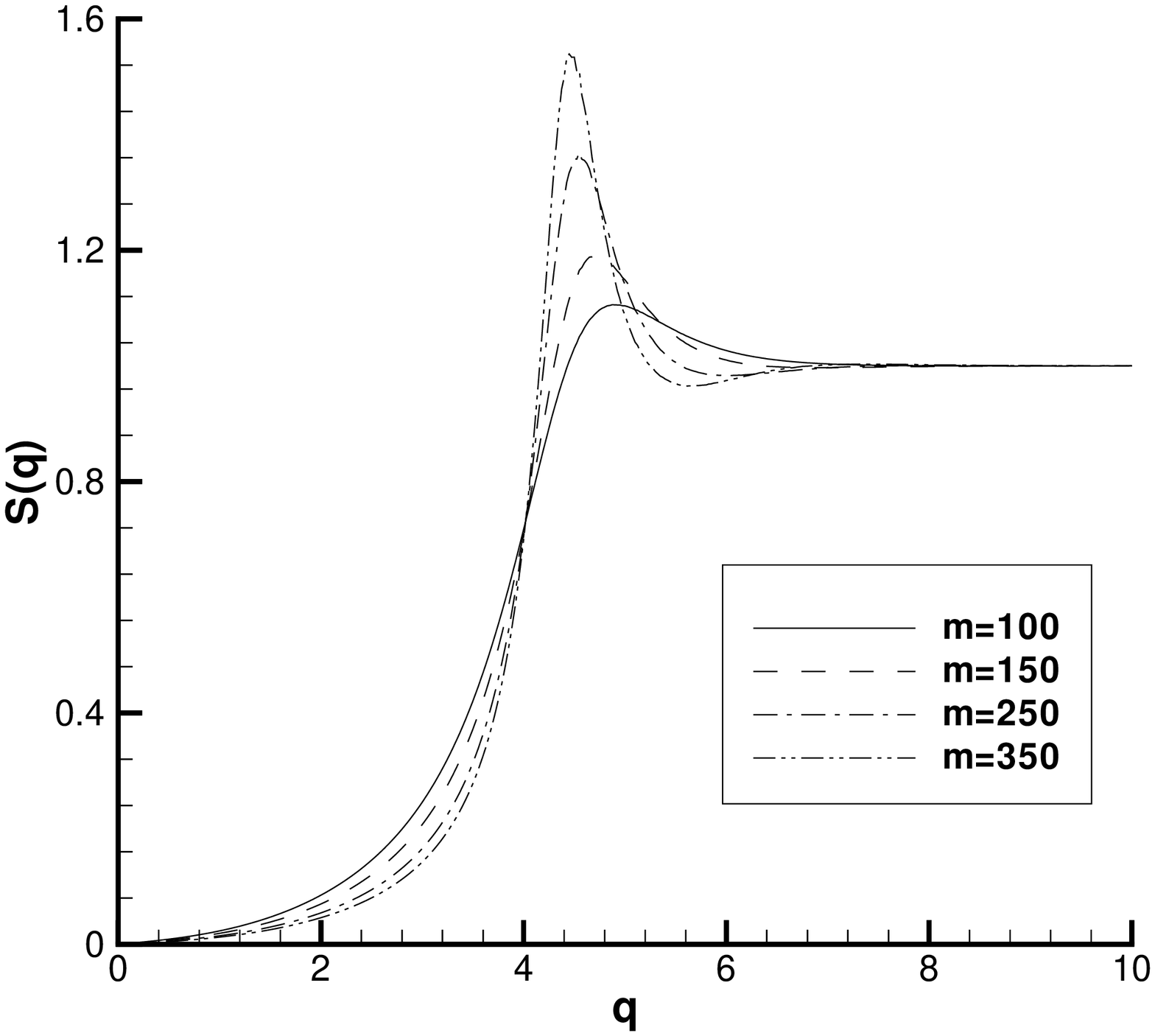}} 
\end{figure}
{\small{
Fig.~1: The static structure factor in terms of $q$, for $r_s=0.6$ 
and different $m$'s.}}

Fig.~1 manifests the very expected behavior of the static structure factor, 
that is the peak of the $S(q)$
is decreased by increasing $T$ (decreasing $m$). 
Disappearing of the peak in the $S(q)$ by increasing the
temperature shows that the system undergoes a phase transition
from the solid phase to the liquid phase. 

We can also see the melting transition 
by fixing the temperature and increasing the magnetic field.
Choosing $m=100$ ($T=50K$), we have found
the $S(q)$ for different values of $r_s$, see Fig.~2. 

\begin{figure}\label{Fig2}
\epsfxsize=8truecm 
\centerline{\epsfbox{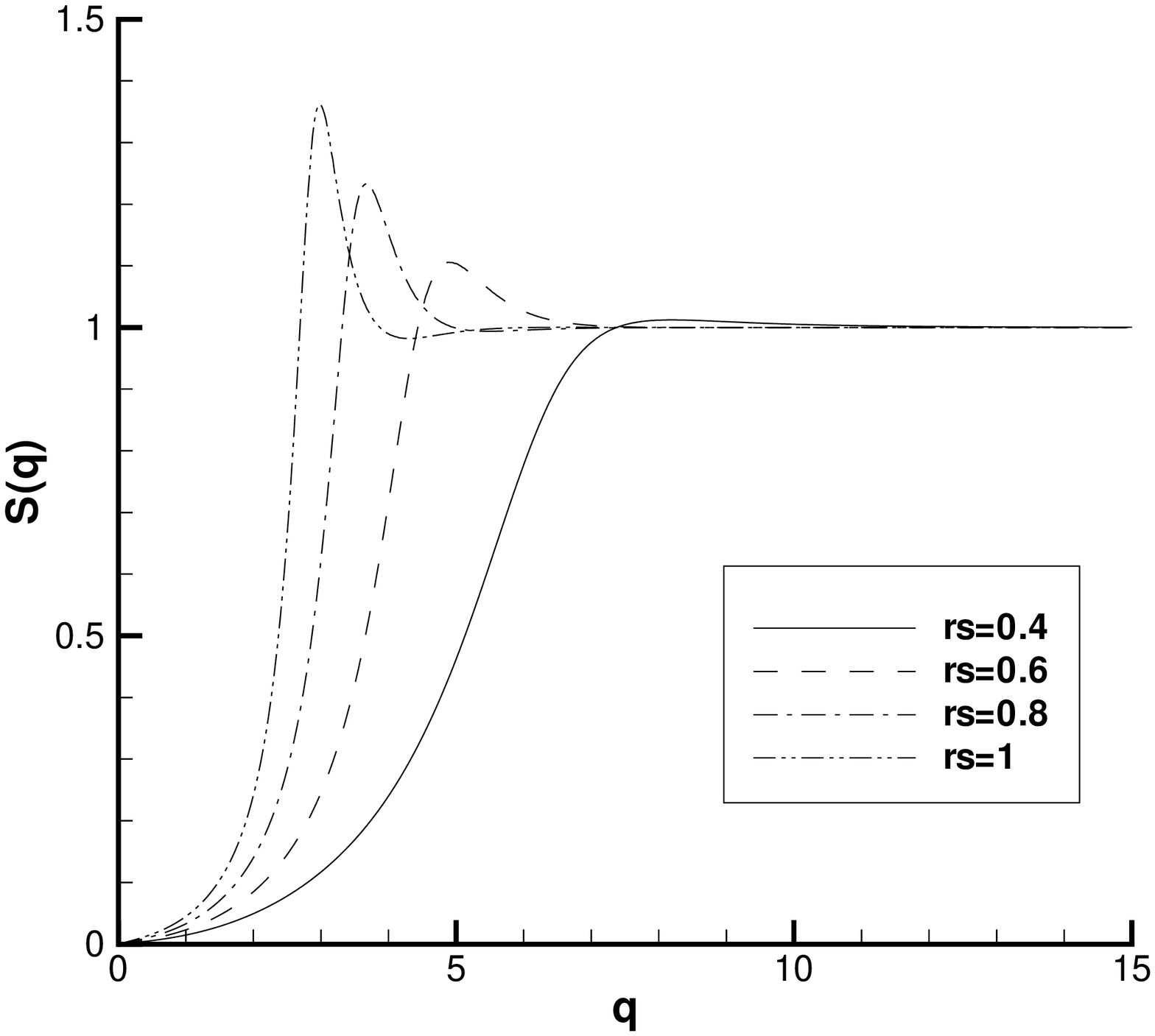}} 
\end{figure}
{\small{Fig.~2: The static structure factor for $m=100$ 
and with different values of $r_s$.} }

The amplitude of the peaks becomes smaller with decreasing the $r_s$,
while it disappears for high magnetic fields. So we will have  
the same scenario for describing the phase transition.

The melting transition of the vortex lattice can also
be discussed by studying the behavior of the pair 
correlation function.
It is seen that the results of $S(q)$ can be used for 
calculating the $g(r)$ defined in
the Eqn.~(\ref{gr}).  
The results of $g(r)$ for $r_s=0.4$ ($B\approx 0.38$(Tesla))
and different masses (temperatures) are plotted in Fig.~3. 

\begin{figure}\label{Fig3}
\epsfxsize=8truecm 
\centerline{\epsfbox{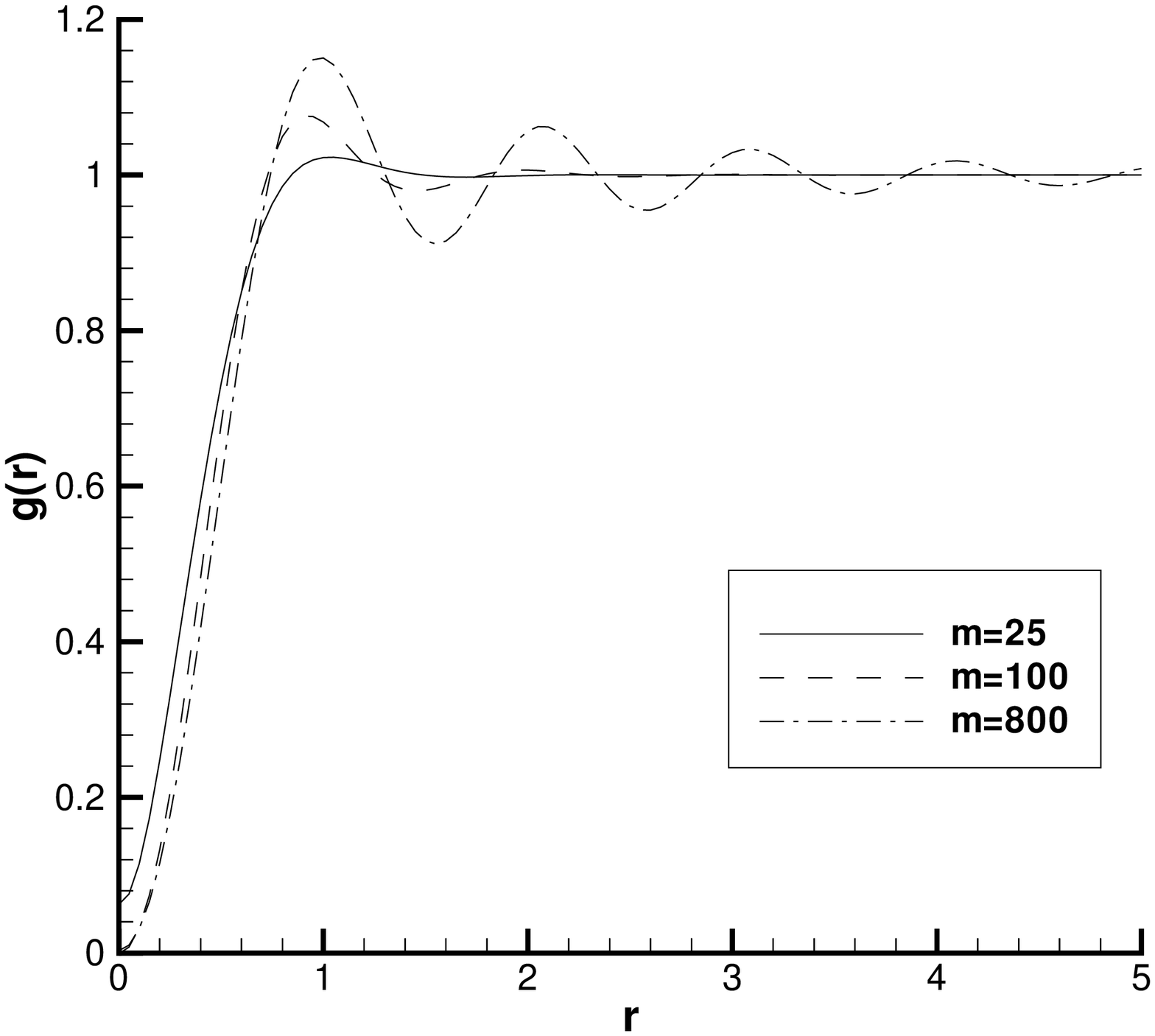}} 
\end{figure}
{\small{Fig.~3: The pair correlation function for $r_s=0.4$ 
with different temperatures.} }

Fig.~3 shows that
the pair correlation function has oscillatory behavior
for low temperatures, however, its
amplitude becomes shorter by increasing the temperature,
and disappears for very high temperatures.
The oscillatory behavior is a signature of the solid
phase, therefore, the system is going to have a transition
from the solid phase to the liquid phase by increasing the temperature.

We have also determined the interaction energy of the system
using the Eqn.~(\ref{Ec}). The results are shown in the Table~1.
$$
\begin{array}{|c|c|c|}\hline
r_s&B&E_{int}\\ \hline
0.1&6&-2.15\\
0.2&1.5&-1.78 \\
0.4&0.38&-1.36 \\
0.6&0.17&-1.09 \\   
0.8&0.09&-0.89 \\
1&0.06&-0.74 \\  \hline
\end{array}
$$               

\vspace{0.2cm}

{\small Table~1: The interaction energy, in STLS
approximation for $m=50$ and different densities (The unit
of magnetic fields is Tesla).}

\vspace{.4cm}

We observe that the numerical results show that the interaction energy 
increases by decreasing the strength of magnetic field.

The results of the pair correlation function and static
structure factor are in good qualitative agreement with 
the expected results of the phase diagram of the vortex system.
However, for obtaining some quantitative description, we use
the spectrum of the excited energy of the system.
We have plotted the $\omega(q)$ in terms of $q$, see Fig.~4. 
Fixing the magnetic field but for varying temperatures,
the following graph resembles that $\omega(q)$
tends to a zero value for a finite $q$ .
The appearance of minimum in $\omega(q)$ is in correspondence with
the fact that $G(q)$ becomes greater than one for some ranges of $m$.
\begin{figure}\label{Fig4}
\epsfxsize=8truecm 
\centerline{\epsfbox{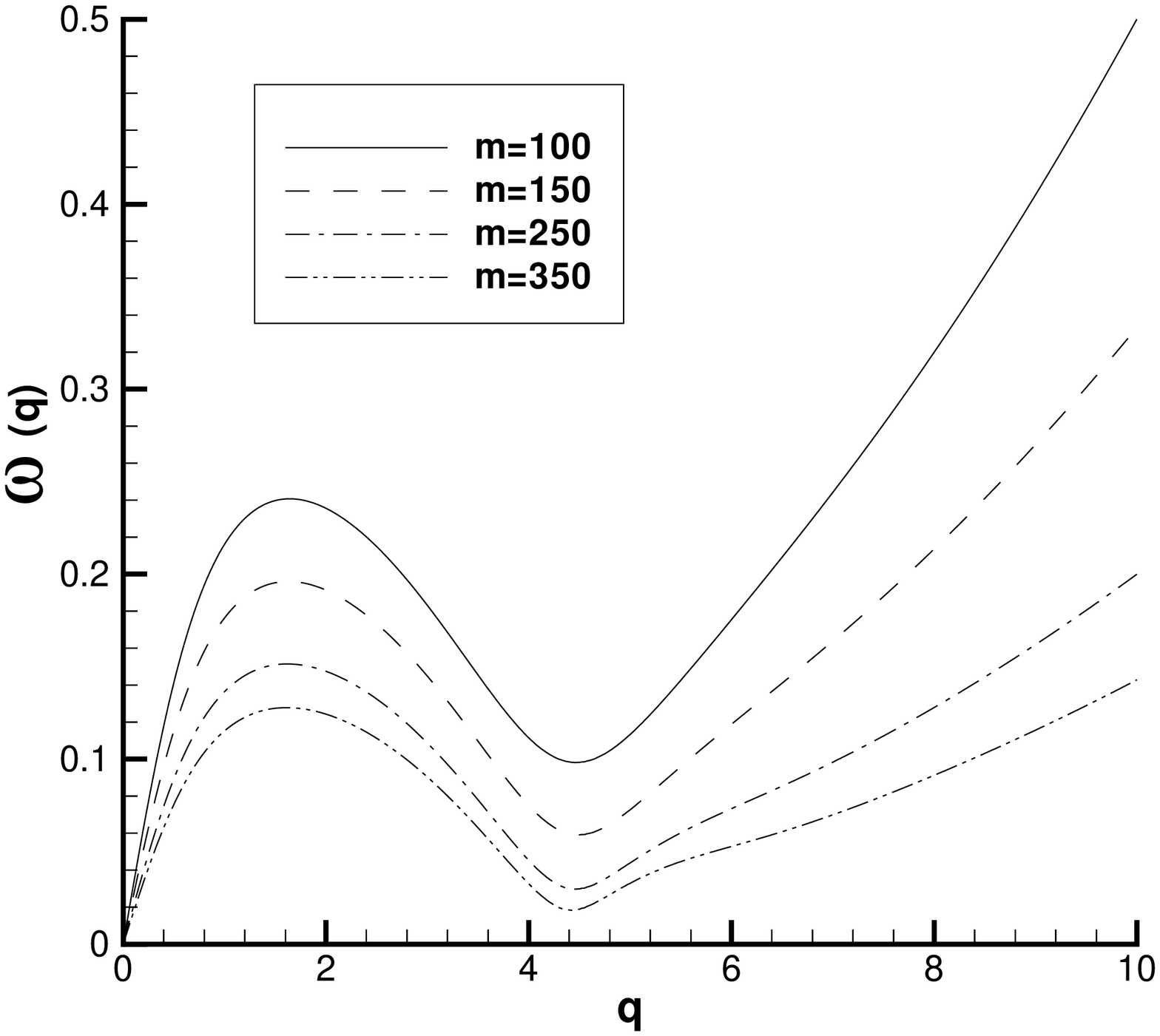}} 
\end{figure}
{\small{Fig.~4: The dispersion relation of the excited energy for
$r_s=0.6$ for different values of $m$.} }

The zero value of the dispersion relation indicate that the
system undergoes a phase transition at that point.
Because of numerical restrictions we are not able to find the 
exact value of the point where $\omega(q)$ becomes zero. However, 
we have realized by our numerical experience that 
when we are close to the transition point the $q$ value
of the minimum $\omega(q)$ and
the local field correction $G(q)$ are not highly sensitive to the
variations of $m$.  
Therefore, at least for determining the transition temperature 
we have ignored their dependence to $m$ 
in the Eqn.~(\ref{omegaq}). 
So we fixed the magnetic field and found the temperature that $\omega(q)$
becomes zero. The resulting $B-T$ phase diagram is plotted in the Fig.~5.

\begin{figure}\label{Fig5}
\epsfxsize=8truecm 
\centerline{\epsfbox{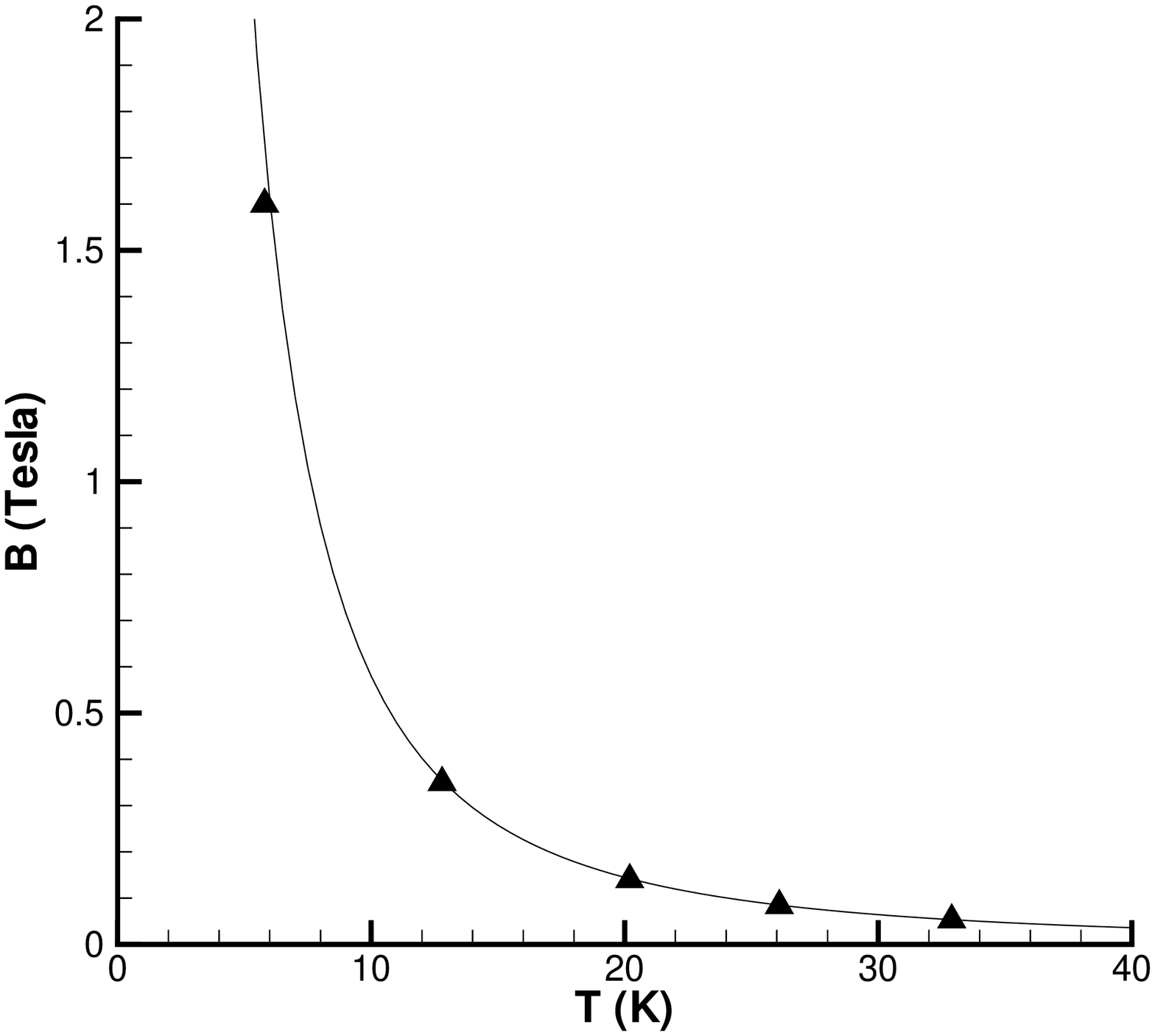}} 
\end{figure}
{\small{Fig.~5: The $B-T$ phase diagram of the flux lines. The solid
line is the predicted result in Ref.\cite{19} produced by Lindemann 
criterion. Triangular points are the results of our numerical 
calculations.} }

Proceeding to fit our data with the known result
$B=4c_L^4\phi_0\varepsilon_0\varepsilon_1/(\sqrt{3}T^2)$ \cite{19,Blatter3},
we found that the best fits can be achieved by fixing 
$c_L\approx 0.15$ and it is observed that the results are supporting 
the reports of Monte Carlo simulations \cite{Blatter4}.

\section{van der Waals interaction}

TAking into account  
the direction dependent interaction for
real vortices give rise to interesting results and predicts
a van der Waals (vdW) interaction 
even for straight vortices \cite{Blatter4}.
Therefore, it would be interesting to consider the
superposition of the short range attractive and the long range
repulsive interactions and study its consequence for the 
low-field phase diagram of the anisotropic superconductors. 
It is shown \cite{Blatter4,Volmer1} that in the decoupled limit, 
$\varepsilon\rightarrow 0$, the interaction potential
is given by
\begin{equation}\label{VWE}
V(R)=v_0\left(K_0(R/\lambda)-a_{vdw}\phi(R/\lambda)
\frac{\lambda^4}{R^4}\right),
\end{equation}
where, $v_0=2\varepsilon_0/T$ measures the amplitude of the
direct interaction between flux lines, and $a_{vdw}$ determines
the strength of the thermal vdW attraction. The function $\phi(x)$ 
smoothly cuts off the power law part for $R <\lambda$ and is defined as,
\begin{equation}\label{phi}
\phi(x) = \left\{
\begin{array}{ll}
0, & x \le x_1 \\
\frac14
\left[ 
1+\sin\left(\pi\frac{x-(x_1+x_2)/2}{x_2-x_1}\right)
\right]^2, 
& x_1<x<x_2 \\   
1, & x \ge x_2
\end{array}
\right.
\label{CutOff}
\end{equation}
with $x_1=1$ and $x_2=5$. The choice of the cutoff function
and the values of $x_1$ and $x_2$ is to some extent arbitrary 
\cite{Blatter4,Volmer1}. The amplitude of the vdW attraction is given
by $a_{vdw}\approx T/(2\varepsilon_0 d \ln^2(\pi\lambda/d))$,
where $d$ is the layer spacing. Using the BSCCO numerical values for the 
parameters, one finds that $a_{vdw}\approx 2\times 10^{-5} T/K$ which is  
$2\times 10^{-3}$ for the temperature $T=100K$ close to critical 
temperature $T_c$ \cite{Volmer1}.

The potential defined in Eqn.~(\ref{VWE}) is applied for calculating the 
static structure factor in the STLS approximation.
The results are plotted in the Fig.~6, where $Avw=a_{vdw}v_0$.  

\begin{figure}\label{Fig6}
\epsfxsize=8truecm 
\centerline{\epsfbox{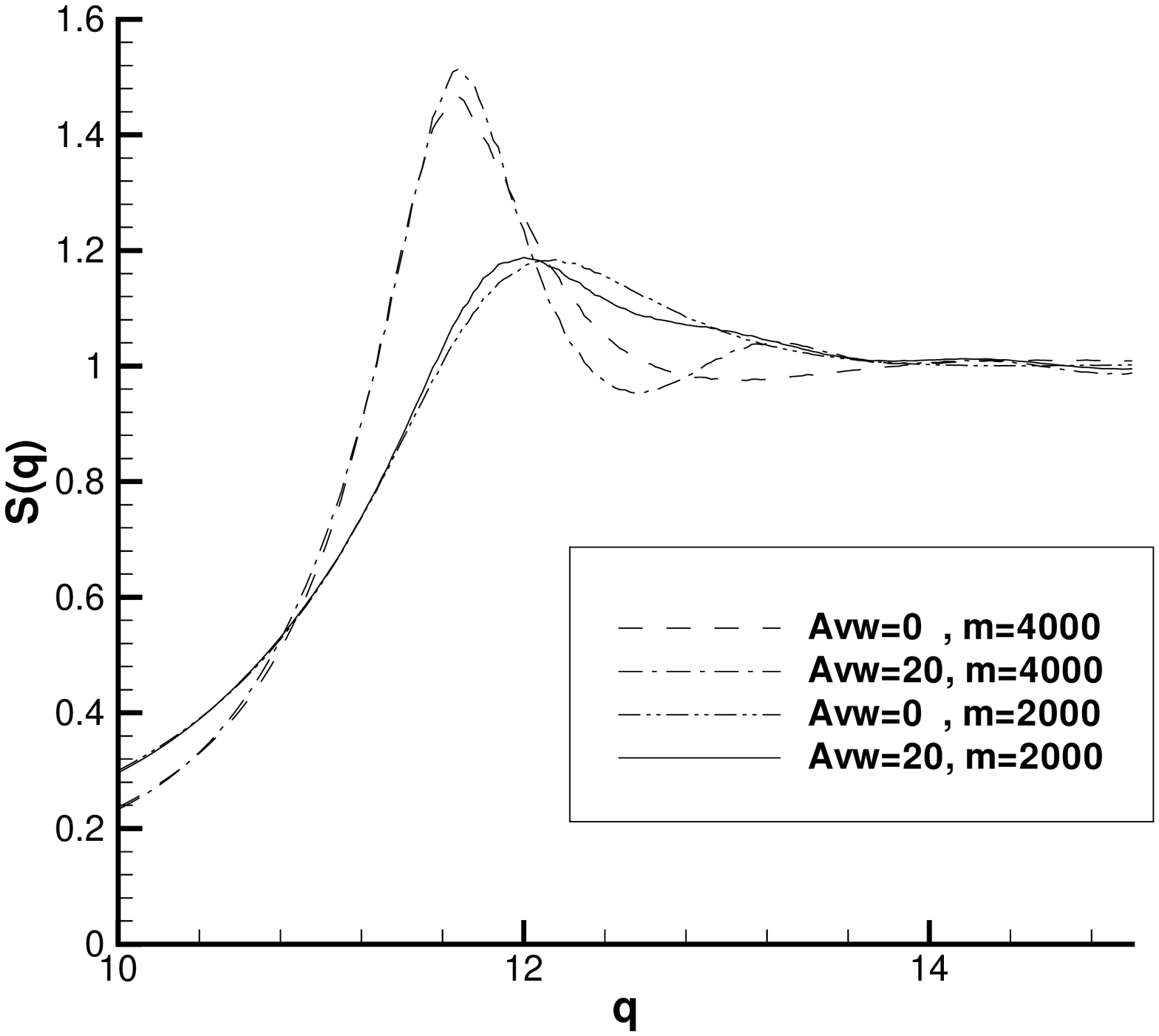}} 
\end{figure}
{\small{Fig.~6: The static structure factor for fixed 
magnetic field $r_s=0.2$ with different temperatures and 
strengths of the vdW attraction. }}

The results show that the importance of the vdW interaction
with high magnetic field is valid only for the temperature close
to the transition. 

One should note that the vdW interaction has very important
consequences for low magnetic fields and temperatures. However,
in our approach it is not possible to consider this case, and 
one has to use some other approximations or Monte Carlo simulations
for seeing these effects.

\section{Conclusions}

In this paper we studied the Bose model in STLS approximation. 
We found the static structure factor, pair correlation function,
interaction energy and spectrum of the excited energies
for different values of the mass (temperature) and the
density (magnetic field).
We discussed that how the results may be applied
for the vortex matter phase diagram. If one fixes the magnetic field 
(temperature)
and calculate $S(q)$ for different temperatures (magnetic field),
the gradual disappearance of the peaks in $S(q)$ manifests 
the existence of the phase transition. 
One can also use the pair correlation function
for describing the melting transition. 
We showed that the changing behavior of $g(r)$ from oscillatory 
to a rather smooth one can be explored within our numerical scheme 
so that the hallmark of the transition from solid 
phase to the liquid phase can be observed . 
The results were in good qualitative agreement with the phase diagram 
of the FLs. For estimation of the $B-T$ phase diagram quantitatively,
we invoke to the behavior of the excited energy spectrum, from which
the resulting 
phase diagram supports the expected results of the high 
temperature superconductors and Monte Carlo simulation quite well.

We also added the van der Waals attractive potential and studied the
effect of both repulsive and attractive potentials in the phase diagram. 
The results indicate that the vdW interaction for high
magnetic field is only important for the temperatures close to the
melting temperature. 
We emphasis that our approach doesn't work for the low magnetic fields
and hence, it would be interesting to use some other methods and
determine the effects of both repulsive and attractive potentials 
in the low magnetic fields and temperatures.

To our knowledge this is the first time that the STLS 
approximation is applied for studying the vortex system,
and our work show that STLS approximation is applicable
for studying other aspects of the vortex systems
and it might help to reveal some unknown properties
in further investigations.

We have benefited from useful discussion with R. Asgari, J. Davoudi
and M. Kardar.
M. Kohandel acknowledges support from the Institute for
Advanced Studies in Basic Sciences, Zanjan, Iran. We
also acknowledge support from Institute for Studies in 
Theoretical Physics and Mathematics, Tehran, Iran.

\end{multicols}{2}
\end{document}